\documentclass[12pt]{iopart}
\usepackage{amssymb,latexsym,mathrsfs}
\usepackage{amsfonts}
\usepackage{graphicx}
\usepackage{color}

\usepackage{color,graphicx}
\usepackage{ulem}

\newcommand{\be}{\begin{equation}}
\newcommand{\ee}{\end{equation}}
\newcommand{\bearr}{\begin{array}}
\newcommand{\enarr}{\end{array}}

\newcommand{\ra}{\rangle}
\newcommand{\la}{\langle}

\def\bea{\begin{eqnarray}}
\def\eea{\end{eqnarray}}
\def\ba{\begin{array}}
\def\ea{\end{array}}

\begin{document}
\title{Matrix Product  States for  Interacting Particles  without Hardcore
Constraints}
\author{ Amit Kumar Chatterjee and P. K. Mohanty}

\address{Condensed Matter Physics Division, Saha Institute of Nuclear Physics, HBNI,
1/AF Bidhan Nagar, Kolkata, 700064 India.}
\ead{amit.chatterjee@saha.ac.in}

\begin{abstract}
We construct  matrix product    steady  state  for a class  of  interacting  particle systems  where 
particles do not obey hardcore exclusion,  meaning each site can occupy any number of particles subjected to the global 
conservation of total number of particles in the system. To represent the arbitrary  occupancy of the  sites,   the matrix 
product ansatz here  requires  an infinite set of matrices  which in turn    leads  to an  algebra   involving  infinite 
number of  matrix equations. We  show that these matrix  equations, in fact,   can be reduced  to a  single  functional 
relation when the matrices are parametric  functions of the representative   occupation number.   We demonstrate  this 
matrix formulation in a class  of stochastic particle hopping processes on a one dimensional periodic lattice where 
hop rates  depend on  the  occupation numbers of   the  departure site and  its   neighbors  
within a   finite range; this includes   some  well known stochastic processes like, totally asymmetric 
zero range process, misanthrope process, finite range process  and  partially asymmetric versions of the same processes  
but with different rate functions depending on the direction of motion. 
\end{abstract}
\maketitle

\section {Introduction}

 Non-equilibrium systems \cite{DDS,book1,book2} are quite common in nature.  Driven   systems    naturally    attain 
 a  non-equilibrium steady-state  (NESS)   with   a nonzero probability  current, in  contrast to equilibrium systems 
 which  satisfy  detailed   balance condition,    ensuring a  zero-current  stationary state  given by  the well 
  known Gibbs  measure \cite{book_db_1,book_db_2}. 
There is no such generic invariant measure  for NESS \cite{DDS,book1};  in general the structure of   NESS  
is  more complex,  yet very interesting,  as they  exhibit  nontrivial  correlations 
\cite{corr_noneq,multi-species,corr_pp_ep}, current reversal  \cite{AZRP_AFRP},  current fluctuations 
\cite{current_fluctuations} and  many other novel phenomena like phase transitions 
\cite{book2,phase_tran_1,phase_tran_2,book3,phase_tran_3,phase_tran_4,phase_tran_5}, even in one dimension.
In absence  of any generic  formalism,  finding   and characterizing   the   stationary measure  of  
non-equilibrium dynamics  is   usually a  non-trivial   task. In spite of these  technical difficulties, 
many exactly solvable models have been found   using special   techniques  like    
Bethe ansatz \cite{bethe1,bethe2,bethe3}, matrix  product ansatz \cite{derrida__tasep_mpa},  
transfer matrix methods \cite{transfer_mat} and methods  of large deviation \cite{ldf} etc.  One of the much 
celebrated examples of solvable   open non-equilibrium  system   is the totally asymmetric simple exclusion 
process (TASEP) \cite{open_TASEP1,open_TASEP2}  where particles   following hardcore repulsion, enter   
to the  system   from left,  move  to  their rightward  vacant neighbor  with unit rate  and  finally exit 
through  the right boundary with   certain rate. 
The steady state of this system  has been obtained, exactly using  matrix product ansatz (MPA) 
\cite{derrida__tasep_mpa} where steady state weight  of any  configuration  is represented  by 
matrix string containing  two non-commuting matrices, one for  occupied site  and the other for  
the vacant site.   Although the dynamics    of TASEP is  quite  simple, it exhibit three  different phases 
and  transitions among them  depending on the  entry and exit rates of particles.  Generalization of  TASEP   
and related models   
\cite{corr_noneq,gen_tasep_1,gen_tasep_2,gen_tasep_3,gen_tasep_4,gen_tasep_5,gen_tasep_6,gen_tasep_7,gen_tasep_8} 
have been very helpful in understanding  NESS  in general.

Matrix product ansatz (MPA)  for  interacting  particle  systems following hardcore constraint  is  known  to be   
one of the most useful and elegant analytical tool for finding NESS. Soon after being    introduced   in context 
of TASEP \cite{derrida__tasep_mpa}, MPA has  found  enormous  applications in different branches of physics. 
MPA \cite{evans_mpa} has been very helpful in calculating spatial correlation functions for exclusion processes 
with point objects \cite{corr_pp_ep}  as well as for extended objects \cite{corr_eo_ep} in one dimension. Study 
of relations between algebraic Bethe ansatz  \cite{algebraic_bethe} and matrix product states for stochastic 
Markovian models in $1-$D \cite{mpa_bethe_markovian_1d} and the same   for spin$-\frac{1}{2}$ Heisenberg 
chains \cite{bethe_mpa_heisenberg_chain} brought calculational convenience and also gave good physical  insight 
to the problems. In connection to correlated non-equilibrium systems, MPA can describe asymptotic distributions 
of the sum of  correlated random variables \cite{mpa_sum_corr_ran_var}. Moreover, as a natural extension of MPA 
on discrete lattice, continuous matrix  product states (cMPS) have been introduced as variational states for $1-$D 
continuum models \cite{cmpa_var_states} and cMPS have already  proved to be convenient in studying Bose gas in $1-$D 
\cite{cmpa_bose_gas} , interacting spin$-\frac{1}{2}$ systems  \cite{cmpa_interacting_fermions} etc. In a nutshell, 
MPA has been attracting interest in vast research areas starting from condensed matter  physics to quantum information 
\cite{mpa_quantum_infor}.
 
 In  matrix product ansatz (MPA), any configuration $\{n_i\}$ ($n_i=0$ or $1$ for exclusion processes) in the 
 configuration space is  represented by  a matrix  string $\{A^{\alpha}_i\}$, where  each matrix $A^{\alpha}_i$ 
 represents either a vacancy $(\alpha=0)$ or a particle of any one of the species $(\alpha=1,2,\dots)$ present 
 at site $i$. Generally the representation of the matrices $A^{\alpha}_i$ do not depend on the  site index $i$. 
 But, notably, particles of different species and vacancies are denoted by matrices that are in general 
 {\it non-commuting} ( $[A^{\alpha},A^{\beta}]\neq 0$ for $\alpha\neq\beta$). If the system is open,   one needs 
 additional   vectors (say $\langle W|$ and $|V \rangle$) to represent  the boundaries.
 The MPA  assumes  that the   steady state   weights of the configuration $\{n_i\}$    to be, 
 
\be P(\{n_i\} \propto  \left\{
\begin{array}{cc} 
  \mathrm{Tr}[\prod_{i=1}^{L} A^{\alpha}_i] &   {\rm periodic} \cr
  \langle W|\prod_{i=1}^{L} A^{\alpha}_i|V\rangle & {\rm open}  .
\end{array}
\right.
\ee
A  specific   stochastic dynamics  on the lattice  insist  the matrices and vectors (if present) to satisfy a set 
of equations, commonly  known as {\it matrix algebra}.  Any representation (of the matrices)  that     satisfy   
this  matrix  algebra  provide  a steady state solution of the   respective dynamics.    
An important point to note is, in all these systems particles   are   constrained  by  {\it hard-core interaction}s, 
that  lead to {\it a finite number} of algebraic equations  to be satisfied.

In this article, we study interacting particle systems   {\it without hardcore interaction},   where each lattice 
site can be occupied by  any number of  particles. To form a matrix product state,  thus,  we  require  infinitely 
many  matrices;   any given dynamics of the system would   then  insist on a   algebra   containing  infinitely many  
matrix equations. It is  not a-priori clear  whether writing the steady state in such a matrix product form is at all 
possible. Here we show that, {\it if the matrices are parametric  function of the occupation numbers},   the   matrix 
algebra,  for  a class of models,  reduce  to a single functional relation  which is easier to deal with. In fact, a  
solution to  {\it this}   functional  relation  eventually leads to an exact steady state   weights  of the model.  
We   demonstrate this in  a class of  interacting particle system where   particles hop  to  one of the nearest neighbor 
with rates  that  depend   on   the    occupation  of the departure site and    its   neighbors  within a  finite range.   

 The  paper is organized  at follows. In  section \ref{sec 2.} we describe a  stochastic process where hop rates are 
 totally asymmetric, i.e.    particles hop only to the rightward neighbor, and   formulate   the matrix product ansatz 
 for the steady state of this model in details. Section \ref{sec 3.} deals with possible matrix product states for some  
 well known models like zero range process (ZRP) \cite{ZRP,ZRPrev}, misanthrope process (MAP)\cite{misanthrope} and finite 
 range process (FRP) \cite{FRP}. Further,  in section \ref{sec 4.}   we study   the  asymmetric particle transfer process 
 where  the  functional form of rate function   for rightward hop is different from    that  of the  leftward  hop.  
 Finally, the summary  of our  results and some discussions  are given  in section \ref{sec 5.}.

\section{Matrix product ansatz in absence of exclusion:}
\label{sec 2.}
In this section we introduce  the  matrix product  formulation  for  interacting  particle systems    in absence of  
hardcore exclusion, i.e. the systems  allow   multiple  occupancy at any lattice site.  We first  consider a generic  
stochastic  process  where particles execute directed motion on a periodic lattice in one dimension ($1$D). The  
dynamics of  the   model is totally asymmetric   in a sense  that  particles  here  hop along a specified  direction 
with hop rates depending  on the  occupancy of several sites, namely the departure site, its   left  neighbors within 
a range   $R_l$ and right neighbors within a range $R_r$.  Below we describe the model in details.

Let the sites  of   the  periodic  lattice be  labeled by $i=1,2,...,L$. With each site $i$, is associated  a 
nonnegative integer variable $n_i (\geqslant 0)$ representing the number of particles at that site (for a vacant 
site $n_i = 0$). The dynamics is as follows. A particle from a randomly chosen site $i$ hops to its right neighbor 
$(i+1)$ with rate $u(n_{i-R_l},..,n_{i-1},n_i,n_{i+1},..,n_{i+R_r}).$ Formally,
\bea
(\dots, n_{i-1}, n_i, n_{i+1},\dots )&&\longrightarrow (\dots, n_{i-1}, n_i-1,
n_{i+1}+1,\dots)  \cr&& 
{\rm with} ~{\mathbf{rate }}~ u(n_{i-R_l}, \dots, n_i, \dots, n_{i+R_r}).~~~ 
\label{eq:dyn1}
\eea 
Clearly,  this driven  non-equilibrium dynamics   conserves the total number of particles $N$ in the system. 
The Master equation dictating the evolution of  probability  $P(\{n_i\})$  of configuration $\{n_i\}$ of the system 
reads  
\bea
 \frac{d}{dt} P(\{n_i\}) &=& \sum_{i=1}^Lu(n_{i-R_l},\dots,n_i+1,n_{i+1}-1,..,n_{i+R_r}) \cr
  && ~~~~~~~  \times  P(\dots,n_{i}+1,n_{i+1}-1,\dots) \cr
  & - & \sum_{i=1}^L  
u(n_{i-R_l},\dots,n_i,\dots,n_{i+R_r})  P(\{ n_i\}).
\label{eq:master}
\eea
In steady state, the net probability flux  must vanish for each configuration $\{n_i\}$, i.e.  the  total  out-flux 
(the first sum on the right hand side of Eq. (\ref{eq:master}))  must balance the  in-flux (the second  sum). 
This cancellation   may  occur  in several different ways, with  detailed balance  being one of the  {\it special} cases 
which, if exists, guarantees  equilibrium. Pairwise balance   is  another   special condition  giving rise to 
non-equilibrium steady states.

For the dynamics in Eq. (\ref{eq:dyn1}), to ensure  that  the  in-flux is balanced by the out-flux 
we first  make an ansatz that the steady state weight $P(\{n_i\})$   can be written in the 
 matrix product form
   \be
  P( \{n_i\}) =  \frac{1}{Q_{L,N}} Tr\left[ \prod_{i=1}^{L}  A(n_i)\right]\, \delta \left(\sum _i n_i -N\right), 
  \label{eq:MPA}
 \ee
where  any configuration is represented by a string of $L$ matrices, $A(n_k)$  being the  matrix  associated with $k-$th 
site containing $n_k$ particles.  The $\delta-$function  here  ensures the particle number conservation and $Q_{L,N}$ is 
the canonical partition function. Now for the ansatz  to be a valid one, we must  ensure that  the matrices  in  
Eq. (\ref{eq:MPA})  satisfy Eq. (\ref{eq:master}) in steady state. This can   be  achieved   by    constructing  a 
suitable   cancellation  scheme  involving  additional  {\it auxiliary matrices}. 
In this context we propose the following cancellation scheme, 
{\small
\bea
\hspace*{-1 cm}
  u(n_{i-R_l},\dots,n_i,n_{i+1}\dots,n_{i+R_r})\quad A(n_{i-R_l})\dots
A(n_{i})A(n_{i+1})\dots A(n_{i+R_r}) \cr \cr
\hspace*{-1 cm}  - u(n_{i-R_l},..n_i+1,n_{i+1}-1..n_{i+R_r}) A(n_{i-R_l}).. A(n_{i}+1)A(n_{i+1}-1).. A(n_{i+R_r})\cr \cr 
 \hspace*{-1 cm} =A(n_{i-R_l})..\widetilde{A}(n_{i})A(n_{i+1})..
A(n_{i+R_r})-A(n_{i-R_l}).. A(n_{i})\widetilde{A}(n_{i+1})..A(n_{i+R_r})\cr 
  \cr \hspace*{-1 cm} =A(n_{i-R_l})\dots
A(n_{i-1}) [\widetilde{A}(n_{i})A(n_{i+1})-A(n_{i})\widetilde{A}(n_{i+1})] A(n_{i+2})\dots A(n_{i+R_r}) 
\label{eq:cancellation_scheme}
\eea
}
where we have introduced a new set of matrices $\widetilde{A}(n)$ -which, in the language of matrix product ansatz, are 
generally known as the {\it auxiliary matrices} \cite{evans_mpa}.
It is straightforward to check that the  above cancellation-scheme  satisfies the Master equation (\ref{eq:master})  
in steady state. What  remains,  is to  find  suitable representation of  the set of matrices  $\{A(n_i)\}$ and the 
{\it auxiliary matrices}  $\{ \widetilde{A}(n_i) \}$   which follow  the   {\it matrix-algebra} given by   
Eq. (\ref{eq:cancellation_scheme}).

A sufficient condition (though not  necessary) to  satisfy Eq. (\ref{eq:cancellation_scheme})  is  
\bea
 \hspace*{-1 cm} u(n_{i-R_l},\dots,n_i,n_{i+1},\dots,n_{i+R_r})\quad A(n_{i-R_l})\dots
A(n_{i})A(n_{i+1})\dots A(n_{i+R_r}) \cr 
   ~~~~~~~~~= A(n_{i-R_l})\dots A(n_{i-1})\widetilde{A}(n_{i})A(n_{i+1})\dots
A(n_{i+R_r})\label{eq:cancellation_scheme_part_one}\\
 \hspace*{-1 cm} u(n_{i-R_l},..n_i+1,n_{i+1}-1..,n_{i+R_r})A(n_{i-R_l})..A(n_{i}+1)A(n_{i+1}-1).. A(n_{i+R_r}) \cr
 ~~~~~~~~~ = A(n_{i-R_l})\dots A(n_{i-1})A(n_{i})\widetilde{A}(n_{i+1})\dots
A(n_{i+R_r}).
\label{eq:cancellation_scheme_part_two}
\eea
 Now both the equations (\ref{eq:cancellation_scheme_part_one}) and (\ref{eq:cancellation_scheme_part_two}) are 
 satisfied consistently if we choose the auxiliary matrix $\widetilde{A}(n)$ to be
 \be
 \widetilde{A}(n)=A(n-1) \quad {\rm for} \quad n\geqq 1 ~~~~~~~~ \mathrm{and} ~~~~~~~ \widetilde{A}(0)=0~. \label{eq:aux}
 \ee
With this choice, Eqs.  (\ref{eq:cancellation_scheme_part_one}) and (\ref{eq:cancellation_scheme_part_two}) reduces   to 
 \bea
  u(n_{i-R_l},\dots,n_i,n_{i+1},\dots,n_{i+R_r})\quad A(n_{i-R_l})\dots
A(n_{i})A(n_{i+1})\dots A(n_{i+R_r}) \cr 
   = A(n_{i-R_l})\dots A(n_{i-1})A(n_{i}-1)A(n_{i+1})\dots A(n_{i+R_r}).
\label{eq:matrix_algebra}
\eea
Thus the matrix product ansatz, formulated here for systems with  multiple   site  occupancy, finally   leads to a  
unique set of equations, as above. For a given   model, with totally asymmetric hop rate $u(.),$  we have to solve the 
matrix algebra (\ref{eq:matrix_algebra}) to find a possible  representation  of    $\{A(0), A(1), \dots \}.$   Practically 
this is a difficult task as  we   need to solve   infinitely many  matrix equations  to be  satisfied   by   infinitely 
large set of matrices $\{A(0), A(1), \dots \}$  which are  non commuting and,  in principle, independent and  unrelated to 
each other. However  for a   generic class of models, which we discuss in  the  following  sections,  it is possible to find 
a matrix representation where $A(n)$  is a  parametric function of  $n;$   i.e.   all  elements of the matrix $A(n)$ are  
specific functions of $n$ i.e. $A(n)_{i,j} = f_{i,j}(n).$   More precisely, the matrices $A(0), A(1), A(2),
\dots$ are not unrelated, rather they are  only  instances  of a general representative matrix  function $A(n).$ In   that 
case,  we  don't  need     to   solve  the matrix  algebra each time separately to obtain  $A(0), A(1), A(2),\dots$,  rather 
we should obtain  the general matrix  function  $A(n)$  by  treating  the  algebra (\ref{eq:matrix_algebra}) as a single 
equation of the matrix function $A(n).$ Once we find   such a  matrix function  $A(n),$ any desired    matrix $A(k)$ can   
be obtained just by putting  the desired value $n=k.$ 
This technique of considering $A(n)$ as a matrix function of the occupation number $n$ and solving the matrix algebra 
just once to find $A(n)$ for general $n,$ is indeed possible for a large class of hop rates which we are going to discuss 
in details in the following sections. Also, since $A(n)$ is now a general matrix function of the site occupation variable 
$n$, we call it    {\it the  site occupation matrix}.

 In order to proceed further, we need to be specific about the dynamics as the matrix algebra $(\ref{eq:matrix_algebra})$  
 explicitly depend on the  hop rates.  In the next section we will consider some stochastic processes like 
 finite range processes (FRP) \cite{FRP} (which   includes as a special case zero range process (ZRP) 
 \cite{ZRP,ZRPrev}), misanthrope process (MAP) \cite{misanthrope} etc.

 \section{ MPA for   Finite  Range Process}
  \label{sec 3.}
  In this section we    consider   finite range process (FRP)  where particles hop   to right   with a hop rate  that  
  depends on  occupation of the departure  site and its  neighbor within a range  $R_l$   to left and $R_r$ to right.  
  In   particular  we discuss  different cases,  $R_l =0= R_r$ (ZRP),      $R_l =0, R_r=1$ (namely misanthrope process),  
  $R_l =1=R_r$ (systems  having pair factorized steady states)  and the generic  scenario with $R_l =R= R_r.$
 For $R_l=R_r=R,$ the hop rate $u(n_{i-R},\dots,n_i,\dots,n_{i+R})$ in FRP depends on { \it same} number $(R)$ of neighbors 
 in   both directions with respect to the departure site $i$.  This special case  was   studied  earlier \cite{FRP} and  it 
 was  shown that,   when the hop-rates  obey certain conditions, FRP leads to  a  $(R+1)$-cluster factorized steady state 
 (CFSS),  $P( \{n_i\}) \sim \prod_{i=1}^L   g(n_i, n_{i+1},\dots n_{i+R}).$  For $R=1$  we  have a  $2$-cluster factorized 
 state,  commonly known  as   the pair factorized state (PFSS) where the steady state weights are 
  $P( \{n_i\}) \sim \prod_{i=1}^L g(n_i,n_i+1).$
  Clearly, a PFSS   can equivalently be written as a matrix product state as  $g(n_i,n_i+1)$ can directly 
  be  considered as  the   elements of an  infinite matrix $T,$  i.e. $T_{n_i,n_{i+1}} = g(n_i,n_{i+1}).$  In the  following 
  we  show that,  whenever a dynamics leads to PFSS,  the  matrix product  ansatz also  naturally  leads to   the same 
  steady state. For 
  $R>1,$  however,  existence of a  cluster factorized state does not  ensure that   it can also be  written  as  matrix 
  product  state. In this section, we show that, even for $R>1,$ one can indeed construct matrix product states through the 
  matrix formulation developed in the previous section.  Below, some examples of totally asymmetric finite range processes 
  for which one obtains steady states in matrix product form, are discussed in details.

  \subsection{Zero range process ($R_l=0=R_r$) } 
  Zero range process (ZRP) \cite{ZRP,ZRPrev} is a very familiar stochastic process  where  hop rate of particles depends 
  only on the  occupation of the  departure site; thus FRP reduces to ZRP  when  $R_l=R_r=0$. One important  feature  of ZRP 
  is that its  steady state  has  a simple factorized form  irrespective of the functional  form of the hop rates, lattice  
  geometry or spatial   dimension.   In spite of having a rather simple dynamics, ZRP shows condensation transition for 
  specific choice of rates - the condensation transition   can be mapped to a  phase separation transition in an  equivalent 
  exclusion process. Interestingly  related   phenomena  like  wealth condensation \cite{wealth_con} in agent based  models, 
  jamming in traffic flow \cite{jamming}   can be related to  condensation transition in ZRP.

  Clearly ZRP  fits into the  generic   matrix product formulation   discussed  in previous section, as the   hop rate  
  $u(n_{i-R_l},\dots,n_i,\dots,n_{i+R_r})$ here is equivalent to $u(n_i).$   Thus for  ZRP,  the matrix algebra   in  Eq. 
   (\ref{eq:matrix_algebra}) reduces to 
  \bea
  u(n_i)\quad A(n_{i-R_l})\dots
A(n_{i})A(n_{i+1})\dots A(n_{i+R_r}) \cr
   = A(n_{i-R_l})\dots A(n_{i-1})A(n_{i}-1)A(n_{i+1})\dots A(n_{i+R_r}),
\label{eq:matrix_algebra_ZRP}
\eea
along with the  auxiliary  matrix   $\widetilde{A}(n)  =  A(n-1),$ as in Eq. (\ref{eq:aux}). 
First we try  for a scalar solution by setting  $A(n)=a(n),$ with $a(n)$ being a positive function for $n\ge 0.$   
This particular choice implies that the auxiliary matrices  for ZRP are  also 
scalar, $\widetilde{A}(n)=a(n-1)$. So Eq. (\ref{eq:matrix_algebra_ZRP}) simplifies to
\bea
  u(n_i) a(n_i)=a(n_i-1) ~~~~
  \Rightarrow ~~~~ a(n)=\frac{a(n-1)}{u(n)}  = a(0) \prod_{k=1}^n  \frac 1 {u(k)} ~.
\label{eq:steady_state_ZRP}
\eea
Thus, the steady state  weight   is 
\be
P(\{n_i\}) \sim  Tr \left[\prod_{i=1}^L A(n_i)\right]  = \prod_{i=1}^L a(n_i)
\ee
 which  is  the   familiar factorized steady state  we know  for ZRP  \cite{ZRPrev}.
 The matrices $A(n)$ being scalar states   that   there is no spatial correlation between occupation at  different  
 lattice sites,  apart from the   global conservation of the total number of particles.

\subsection{Misanthrope Process ($R_l=0 , R_r=1$):}
 Misanthrope process \cite{misanthrope}  
is a special case  of FRP    with  $R_l=0$ and $R_r=1,$ i.e., the hop rate $u(n_i,n_{i+1})$ here  is no 
longer departure-site symmetric,   
it depends on the occupation number of the departure site $i$ and the arrival site $(i+1)$ only.  For certain choice of hop 
rates, misanthrope process  is known to have  a factorized steady state as studied in \cite{misanthrope,misanthrope_fss}. 
Here we will show that the same factorized state can be obtained starting from  matrix product ansatz.  Note, that  the   
generic choice of auxiliary matrices  $\widetilde{A}(n)=a(n-1)$ along  with $A(n)=a(n)$, given by  equations (\ref{eq:aux})  
and (\ref{eq:matrix_algebra}),  would  result in  $u(n_i, n_{i+1})=\frac{a(n_i-1)}{a(n_i)}$ which   is inconsistent as  this 
choice   does not allow the  hop  rate  to  depend  on the occupation of  the  arrival site.  We now proceed  with  a  
scalar choice $A(n)  = a(n), \widetilde{A}(n)=\widetilde a(n),$ where    both functions   $a(n)$ and  $\widetilde a(n)$ 
are yet to be determined. The  cancellation scheme in Eq. (\ref{eq:cancellation_scheme})   now  becomes
\be
 u(n_i,n_{i+1})-u(n_i+1,n_{i+1}-1)\frac{a(n_i+1)a(n_{i+1}-1)}{a(n_i)a(n_{i+1})} =\frac{\widetilde{a}(n_i)}{a(n_i)}-
\frac{\widetilde{a}(n_{i+1})}{a(n_{i+1})}.
\label{eq:matrix_algebra_Misanthrope}
\ee
Since the hop rates $u(m,n)=0$  for $m<1$  or for $n<0,$   the above equation,  for 
$n_{i+1}=0$ and for $n_i=0$  reduces to, 
\be
u(n,0)= \frac{\widetilde{a}(n)}{a(n)}\,-\,\frac{\widetilde{a}(0)}{a(0)}=u(1,n-1)\frac{a(1)}{a(0)}\frac{a(n-1)}{a(n)}.
\label{compare_MAP}
\ee
These equations  further  results in, 

\bea
a(n)=\,\frac{a(1)}{a(0)}\frac{u(1,n-1)}{u(n,0)} a(n-1)=a(0)\left(\frac{a(1)}{a(0)}\right)^{n}\,\prod_{k=1}^{n}\frac{u(1,k-1)}{u(k,0)}
\label{FSS_MAP}\\
\widetilde{a}(n)=\left[ \frac{\widetilde{a}(0)}{a(0)}+u(n,0) \right] a(n).
\label{eq:auxiliary_Misanthrope}
\eea
It appears   from   Eq. (\ref{FSS_MAP}) that  we have  a  factorized steady state    
$P(\{n_i\}) \propto \prod_{i=1}^L a(n_i)$ for any hop rate  $u(m,n)$ which is  certainly not true, because  these  
equations (\ref{compare_MAP}) and (\ref{eq:auxiliary_Misanthrope}), derived by  using specific boundary conditions 
($n_i=0$ or $n_{i+1}=0$), must  also  respect   Eq. (\ref{eq:matrix_algebra_Misanthrope})   for all $n_i>0, n_{i+1}>0.$ 
Using Eq. (\ref{eq:auxiliary_Misanthrope}) in Eq.(\ref{eq:matrix_algebra_Misanthrope})  we get 
\bea
 \hspace*{-1.5 cm} u(n_i,n_{i+1})-u(n_i+1,n_{i+1}-1)\frac{a(n_i+1)a(n_{i+1}-1)}{a(n_i)a(n_{i+1})} = u(n_i,0)-u(n_{i+1},0).
\label{eq:map_condition}
\eea
Thus,  in MAP,   we  have a factorized steady  state  $P(\{n_i\}) \propto \prod_{i=1}^L a(n_i)$  only when  the  hop  
rate  $u(n_i,n_{i+1})$ satisfy  Eq. (\ref{eq:map_condition}), which  is  the  familiar constraint that  has  been   
reported  earlier \cite{misanthrope_fss}. The steady state  weights given by  Eq. (\ref{FSS_MAP})   is also  identical 
to   the  one  which is already known for MAP \cite{misanthrope_fss}.

  \subsection{FRP with $R_l=R_r=R=1$}    
 If $R_l=R_r=R=1$, particle from an occupied site $i$ hops to site $(i+1)$ with rate $u(n_{i-1},n_i,n_{i+1})$ that depends 
 on the occupancies of the departure site $i$, its {\it left nearest} and {\it right nearest} neighbors ($(i-1)$ and 
 $(i+1)$ respectively). For a special class of hop rates, this  finite range process has a pair factorized steady state 
 (PFSS) \cite{pfss} given by
 \be
P(\{n_i\})  = \frac{1}{ Z_{L,N}} \prod_{i=1}^L   g(n_i, n_{i+1})  \delta( \sum_i n_i -N). \label{eq:PFSS}
\ee
Obviously   $g(n_i,n_{i+1})$ itself  can be  considered as elements  of  an infinite dimensional matrix, infinite 
dimensional because  $n_i$ and $n_{i+1}$  can take arbitrarily  large positive  integer values. In other  words, PFSS is a  
matrix product state represented by infinite dimensional matrices.   We   show  below that  for    a class  of models   
one can  obtain finite dimensional representation.  

Let us consider hop rates of the form
\be
u(n_{i-1},n_i,n_{i+1})=\frac{\langle \alpha(n_{i-1})\mid \beta (n_i-1) \rangle \langle 
\alpha(n_i-1)\mid \beta (n_{i+1}) \rangle }{\langle \alpha(n_{i-1})\mid \beta (n_i) \rangle \langle 
\alpha(n_i)\mid \beta (n_{i+1}) \rangle} ,
\label{eq:rate_PFSS}
\ee
$\alpha(n)$ and $\beta(n)$ are arbitrary positive functions with $\alpha(-1)=\beta(-1)=0$. 
It  is easy to see that  the   matrix algebra (\ref{eq:matrix_algebra}), along with  the   choice of   auxiliary 
$\widetilde{A}(n)=A(n-1)$ as in Eq.  (\ref{eq:aux}),  can be  satisfied   if, 
\be
A(n)=|\beta(n)\rangle \langle \alpha(n)|.
\label{eq:matrices_nonequilibrium_R=1}
\ee
Correspondingly, the steady  state probability  of configurations are  $ P(\{n_i\}) \sim  
Tr\left[\prod_i  A(n_i)\right] \delta (\sum _i n_i -N).$ The grand canonical  partition function is then 
\be 
\hspace*{-1.0 cm}Z_L(z)= \sum_{ \{n_i\}} z^{n_i} Tr \left[\prod_{i}^L  A(n_i)\right] = Tr\left[ T(z) ^L\right] ;~~  
T (z) = \sum_n z^n |\beta(n) \ra  \la \alpha(n)|.
\ee
Now  one can  conveniently calculate   steady state average  of desired observables  
in the steady state, like spatial correlations, density fluctuations, particle current etc.. 
For example, since the particle hops only towards right,   the average steady state  current  of the system  is
\be
J =   \la  u(n_{i-1},n_i,n_{i+1}) \ra  = \frac{ 1}{Z_L(z)}  
\sum_{ \{n_i\}}  u(n_1,n_2,n_3) z^{n_i} Tr \left[ \prod_{i}^L  A(n_i)\right] = z.
\ee
To   find the dependence of $J$   on  the average particle density $\rho,$   one must  calculate  
$\rho(z) = \frac{z}{Z_L} \frac{d}{dz} Z_L$   and  then  invert this  relation.

\subsection{Finite range process with $R_l=R_r=R>1$} 
 For a more  general finite range process (FRP) corresponding to $R_l=R_r=R>1$ the hop rate 
 $u(n_{i-R},\dots,n_i,\dots,n_{i+R})$ is a function of  $(2R+1)$ site variables, namely the occupation number of the 
 departure site and that of $R$ neighbors  to its left and to right.  This model was introduced earlier in Ref. \cite{FRP} 
 where  it has been  shown that the  steady state   of the system   is cluster factorized when the hop rates $u(.)$  satisfy 
 certain  specific  conditions. For a  cluster factorized steady state (CFSS), the  probability of configurations are given 
 by,   
 \be
P(\{n_i\})  = \frac{1}{ Z_{L,N}} \prod_{i=1}^L   g(n_i, n_{i+1},\dots,n_{i+R})  \delta( \sum_i n_i -N), \label{eq:CFSS}
\ee
where $ g(n_i, n_{i+1},\dots,n_{i+R})$, a function of $(R+1)$ variables, is known as the cluster weight function, and 
$Z_{L,N}$ is the   canonical partition function.  
The authors   in \cite{FRP}   have   restricted  their study to  FRP  where the cluster weight function  has  a `sum-form'  
$ g(n_i, n_{i+1},\dots,n_{i+R}) =\sum_{k=0}^{R} f_k(n_{i+k}).$
For example, when $R=2$, FRP has  a 3-cluster factorized steady state   with weight function  
$g(n_i, n_{i+1},n_{i+2})=\gamma_0(n_i)+\gamma_1(n_{i+1})+\gamma_2(n_{i+2})$   if the  hop rate   (that  satisfies 
the required condition) is 
\bea \hspace*{-1cm}
u(n_{i-2},n_{i-1},n_i,n_{i+1},n_{i+2})=\prod_{k=0}^{2} \frac{\gamma_0(n_{i-2+k})+\gamma_1(n_{i-1+k})+
\gamma_2(n_{i+k}-1)}{\gamma_0(n_{i-2+k})+\gamma_1(n_{i-1+k})+\gamma_2(n_{i+k})}.
\label{eq:PFSS_sumform}
\eea
Clearly $g(.)$   being a  function of $(R+1)$ variables,  unlike for $R=1$ case,   it can not  be   considered  directly as  
matrix when $R>1$. Thus, for $R>1$, rewriting   a cluster factorized  steady  state  as   a  matrix product state is  already 
challenging. Moreover, here  we  will discuss more generalized forms of the hop rates which does not necessarily  lead to 
the  `sum-form' of the cluster weight function.  Let us consider an example  $R_l=R_r=R=2,$  where  a particle from a 
randomly chosen site $i$ hops to its right  neighbor $(i+1)$ with a rate 
\bea\hspace*{-2.5 cm}
 u(n_{i-2},n_{i-1},n_i,n_{i+1},n_{i+2})= \prod_{k=0}^{2}
\frac{\langle f_0(n_{i-2+k}) |f_1(n_{i-1+k}) \rangle +\langle f_2(n_{i-1+k})|f_3(n_{i+k}-1) \rangle}
{\langle f_0(n_{i-2+k}) |f_1(n_{i-1+k}) \rangle +\langle f_2(n_{i-1+k})|f_3(n_{i+k}) \rangle}
\label{eq:rate_R=2}
\eea
where  $\la f_\nu(n) |= ( h_\nu^1(n), h_\nu^2(n),  h_\nu^2(n), \dots h_\nu^d(n))$   are    $d$-dimensional row-vectors 
and  $ |f_\nu(n) \ra = \la f_\nu(n)|^T$ (here $\nu=0,1,2,3$).  
In fact the rates  here satisfy the  conditions required \cite{FRP} for a  system to have  3-cluster factorized steady state
$P(\{n_i\})  \sim  \prod_i   g(n_i, n_{i+1}, n_{i+2})$  with 
\be
g(l,m,n)= \langle f_0(l) |f_1(m) \rangle +\langle f_2(m)|f_3(n)\ra. \label{eq:g_R=2}
\ee
Although we have  exact steady state weights for these  rates, it is  not  very useful in calculating the partition function 
or other physical observables. This is because,   any    occupation  variable $n_i$ appears   thrice    in  the   cluster 
factorized   state  and carrying  out  the   sum over the all possible values of $n_i$  is not straightforward.   In this 
regards, the matrix formulation, where  the matrices are parametrized by the {\it local} occupation  number, is    very 
helpful. In the following  we   proceed with the  MPA and  use  the  auxiliary matrices $\widetilde{A}(n)=A(n-1),$ as 
in Eq. (\ref{eq:aux}). The matrices    $A(n)$ should then    follow   the  matrix algebra  given by  
Eq. (\ref{eq:matrix_algebra}) with hop rate there  replaced by   Eq.  (\ref{eq:rate_R=2}).   We find that this algebra  
is  satisfied  by  the following representation of matrices,
\be
A(n) = \left( | \beta  (n) \ra \otimes I\right)  \Gamma(n)  \left(  I \otimes \la \alpha(n) | \right)   \label{eq:matrices_nonequilibrium_R>1}
\ee
where, 
\bea
| \beta  (n) \ra =  \left(\begin{array}{c}   1 \\ | f_3  (n) \ra  \end{array}\right);~~~
\la \alpha(n) |  =  \left(\begin{array}{cc}\la f_0(n)| & 1 \end{array}\right)
\eea
 are $(d+1)$-dimensional vectors and 
 \be 
 \Gamma(n) =  \left(\begin{array}{cc}  | f_1  (n) \ra & 0_{d\times d}\\  0 & \la f_2(n)|   \end{array}\right)
 \ee 
 is a $(d+1)$-dimensional matrix. Also, $I$  is the identity matrix  in  $(d+1)$  dimension. The operation $\otimes$ is the familiar {\it direct product}. 
 Note that    $ |\beta  (n) \ra \otimes I $  and  $I \otimes \la \alpha(n) |$ are not square matrices; their dimensions 
 are respectively  $(d+1)^2 \times (d+1)$  and  $(d+1) \times (d+1)^2.$
Thus, the   dimension of the  matrices   $A(n)$   that  represent the  steady state  weights is $(d+1)^2.$ 
In the Appendix we have discussed  how to  generate  the matrix representation  
systematically   for    a dynamics  (\ref{eq:rate_R=2})  or equivalently   for  a  model which has a cluster factorized 
steady  state with cluster weight function $g(.)$   given by   Eq.  (\ref{eq:g_R=2}).

Let us  illustrate the  dynamics and the steady state  weights   for a specific  example  where  the hop rates are 
given by  Eq. (\ref{eq:rate_R=2})  with  scalar choice of  $\la f_\nu(n) |,$    i.e, 
$\la f_\nu(n) | =f_\nu(n) = |f_\nu(n)\ra.$  Explicitly, the hop rates  are now 
\bea
\hspace*{-2.2 cm} u(n_{i-2},n_{i-1},n_i,n_{i+1},n_{i+2})= \prod_{k=0}^{2}
\frac{ f_0(n_{i-2+k})~ f_1(n_{i-1+k}) + f_2(n_{i-1+k})~f_3(n_{i+k}-1) }
{ f_0(n_{i-2+k}) ~f_1(n_{i-1+k})  + f_2(n_{i-1+k})~f_3(n_{i+k}) }.
\label{eq:5site_dyn}
\eea
For this simple choice of hop rate, 
\bea \hspace*{-1 cm}
| \beta  (n) \ra =  \left(\begin{array}{c}   1 \\  f_3  (n)  \end{array}\right);~
\la \alpha(n)|   =  \left(\begin{array}{cc} f_0(n) & 1 \end{array}\right);~
\Gamma(n) =  \left(\begin{array}{cc}   f_1  (n)  & 0\\  0 &  f_2(n)   \end{array}\right),\nonumber
\eea
and correspondingly  the steady state matrix $A(n),$ from Eq. (\ref{eq:matrices_nonequilibrium_R>1}), reduces to a 
$4-$dimensional matrix  
\bea
\hspace*{-1.0 cm} A(n)=\left( \begin{array}{cccc}
f_0(n)f_1(n) &  f_1(n) & 0 & 0 \\
0 & 0 & f_0(n)f_2(n) & f_2(n) \\
f_0(n)f_1(n)f_3(n) &  f_1(n)f_3(n) & 0 & 0 \\
0 & 0 & f_0(n)f_2(n)f_3(n) & f_2(n)f_3(n) \\
             \end{array}
\right).
\nonumber
\eea
Thus, we   obtain   the matrix  product steady state $P(\{n_i\}) \sim  Tr [\prod_{i=1}^L A(n_i)]$ for the dynamics 
(\ref{eq:5site_dyn}).  As we have  already mentioned,   the  steady state  of this dynamics  has  3-cluster factorized form
$P(\{n_i\})  \sim  \prod_i   g(n_i, n_{i+1}, n_{i+2}).$   
Finally,  once  the representation of   matrices  $A(n)$ as in (\ref{eq:matrices_nonequilibrium_R>1})  are   known,   
it is quite straight forward to calculate the partition function and  any desired  observable. 
\section{ MPA for finite range process  with asymmetric  rate functions:}\label{sec 4.} 

 In the previous sections, we have studied    finite range processes where particles hop   only to the  right.  In   fact,
 the  steady state measure   of   FRP     remains   invariant  if  we  introduce  a  parameter  $p,$  the probability   that a  particle  
 chooses  the right neighbor as  a target site  and moves   there  with rate $u(.)$  or with  probability  $(1-p)$ it decides to hop to 
 left and moves there with the same rate $u(.)$.  A non-trivial situation is when  the functional form of  rate 
functions for right hop  is different   from  that of the left hop. A class of such asymmetric motion of 
particles without hardcore constraints has recently been introduced and studied in \cite{AZRP_AFRP} in context of asymmetric zero range 
process (AZRP), asymmetric misanthrope process (AMAP) and asymmetric finite range process (AFRP); each of them having exactly solvable 
non-equilibrium invariant measures -factorized steady states (FSS) for AZRP, AMAP and cluster factorized steady states (CFSS)  for AFRP. 
AZRP, AMAP show interesting features like  density dependent current reversal (keeping the external bias fixed), condensation (tuned by 
the proportion of right and left moves executed by the particles)- phenomenas solely induced by different functional forms of the  left 
and right rates. Now AZRP and AMAP, having FSS,  would not be of much interest in context of matrix product states since in the previous 
sections  we have already discussed  how  the matrices and  the  auxiliaries reduce to scalars  for  a steady state  to   have a factorized 
form.  So we would like to explore  only the  possibility of obtaining a   matrix product state for  AFRP.  
In this section we will first introduce a very general dynamics for asymmetric hopping process in one dimension which includes AFRP as a 
special case. We will then illustrate  the matrix formulation with   some examples. 

\subsection{General asymmetric   hopping dynamics }
Let us consider an interacting particle system on a one dimensional periodic lattice where particles (without hardcore exclusion) can hop 
in both directions with respective forward and backward rates; the  rate  functions  depend on the occupation of several lattice sites as 
well as on the direction of motion of the particles, i.e., the right and left hop rates can have different functional forms. The model  is 
defined on a  one dimensional periodic lattice with $L$ sites where each site $i$ contains $n_i$ particles with $n_i (\geqslant 0)$ being a 
nonnegative integer. A particle from site $i$ (with $n_i>0$), 
can  move either to its immediate right neighbor $(i+1)$ with rate $u_R(n_{i-R_l},\dots,n_i,\dots,n_{i+R_r})$ or it can hop to its 
immediate left neighbor $(i-1)$ with rate $u_L(n_{i-R'_l},\dots,n_{i},\dots,n_{i+R'_r})$. Note that,  the model is different from 
the  one discussed in Ref. \cite{AZRP_AFRP} as   not only the forward and backward rates 
have different functional forms $u_R(.)$ and $u_L(.)$, also, they have  different number of arguments;  the right hop rate depends on $R_l$ 
left neighbors and $R_r$ right neighbors   in contrast to $R'_l$ left and $R'_r$ right neighbors for the left hop rate. In general, all four 
numbers $R_l, R_r, R'_l, R'_r$ can be different. We ask if this stochastic process can lead to a non-equilibrium steady state, 
in matrix product form. Solving the matrix algebra to find out a matrix product state for arbitrary values of $R_l,R_r,R'_l,R'_r$ appears  to be 
quite complex. We  restrict our selves to some  special cases and study below two specific examples.\\

\subsubsection{Example 1 :}
 Our first example is $R_l=1\neq R'_l=2$ and $R_r=2\neq R'_r=0$, i.e., a particle from site $i$ hops  to  the  right neighbor with rate  
 $u_R(n_{i-1},n_i,n_{i+1},n_{i+2})$  and  it hops to the left with rate   $u_L(n_{i-2},n_{i-1},n_i).$  This  dynamics  has not been  
 studied  earlier  in context of particle or mass transfer processes  and clearly  the  criteria for having a  factorized   or  
 cluster-factorized steady state  is   not    known. In   the following    we show, using a   specific example, that   one can use   MPA   
 to  obtain  an exact steady state weights  of these models in  some special cases.

 Let  us   choose the    rate functions in the following form 
 \bea
 u_R(n_{i-1},n_i,n_{i+1},n_{i+2})&=&u(n_{i-1},n_i,n_{i+1})\,+\,v(n_i,n_{i+1},n_{i+2}) \cr ~~~~~~~u_L(n_{i-2},n_{i-1},n_i)&=&v(n_{i-2},
 n_{i-1},n_i). 
 \label{eq:asym_diff_1}
 \eea
 Here,  the right hop rate $u_R(.)$  is  a sum of two independent  functions- the first part $u(.)$ is symmetric with respect to the 
 departure site $i$ and the  rest  $v(.)$ is symmetric about the arrival site $(i+1)$. On the other hand, the left hop rate 
 $u_L(.)\equiv v(.)$ is symmetric with respect to the arrival site. Assuming  that   the steady state  of  the  model can be written in a 
 matrix product  form  $P( \{n_i\})  \sim  Tr \left[\prod_i  A(n_i)\right] \delta (\sum _i n_i -N),$ the  
 Master equation for  dynamics (\ref{eq:asym_diff_1})   in  steady state reduces to, 
 \bea
  \sum_{i=1}^L  [u(n_{i-1},n_{i},n_{i+1})+ v(n_{i},n_{i+1},n_{i+2})+
  v(n_{i-2},n_{i-1},n_{i})]\cr Tr[\dots
A(n_{i-2})A(n_{i-1})A(n_{i})A(n_{i+1})A(n_{i+2})\dots ] \cr
  -\sum_{i=1}^L[ u(n_{i-2},n_{i-1}+1,n_{i}-1) Tr[.. A(n_{i-2})A(n_{i-1}+1)A(n_{i}-1)..]\cr
  +v(n_{i-1}+1,n_{i}-1,n_{i+1}) Tr[.. A(n_{i-1}+1)A(n_{i}-1)A(n_{i+1}).. ]\cr\cr
  +v(n_{i-1},n_{i}-1,n_{i+1}+1) Tr[.. A(n_{i-1})A(n_{i}-1)A(n_{i+1}+1).. ]]=0~.
\eea
The above equation can be equivalently written as
\be \sum_{i=1}^{L} \mathrm{Tr}[ \dots A(n_{i-2})\mathbf{F}(n_{i-1},n_{i},n_{i+1}) A(n_{i+2})\dots ]\,=\,0, \label{eq:ex1}\ee
where, 
{\small 
\bea
\hspace*{-2.5 cm} \mathbf{F}(n_{i-1},n_{i},n_{i+1})\,=\, \cr \hspace*{-2.5 cm}[u(n_{i-1},n_{i},n_{i+1})A(n_{i-1})A(n_{i})A(n_{i+1})-u(n_{i-1},n_{i}+1,n_{i+1}-1) A(n_{i-1})A(n_{i}+1)A(n_{i+1}-1)]\cr
\hspace*{-2.5 cm} +[v(n_{i-1},n_{i},n_{i+1}) A(n_{i-1})A(n_{i})A(n_{i+1})-v(n_{i-1}+1,n_{i}-1,n_{i+1}) A(n_{i-1}+1)A(n_{i}-1)A(n_{i+1})]\cr
\hspace*{-2.5 cm} +[v(n_{i-1},n_{i},n_{i+1}) A(n_{i-1})A(n_{i})A(n_{i+1})-v(n_{i-1},n_{i}-1,n_{i+1}+1) A(n_{i-1})A(n_{i}-1)A(n_{i+1}+1)].\nonumber
\eea
}\\
Equation  (\ref{eq:ex1})  is a  sum of $L$  similar terms where each term carries a three site function 
$\mathbf{F}(x,y,z)$ that contains  the relevant information about the dynamics, i.e.,  the in-flux and out-flux for a given configuration. 
So it would be reasonable to find a local three site cancellation scheme for $F(x,y,z)$  that would make  the 
sum of $L$ terms  in Eq.  (\ref{eq:ex1})  equal to zero. 
We  propose  the  following cancellation scheme,    
\bea
  \mathbf{F}(n_{i-1},n_{i},n_{i+1}) \, =\,[A(n_{i-1})\widetilde{A}(n_{i})A(n_{i+1})-A(n_{i-1})A(n_{i})\widetilde{A}(n_{i+1})]\cr
\qquad \qquad \qquad \,\,\,\,\, +\,\,\,[\widehat{A}(n_{i-1})\bar{A}(n_{i})A(n_{i+1})-A(n_{i-1})\widehat{A}(n_{i})\bar{A}(n_{i+1})]\cr
\qquad \qquad \qquad \,\,\,\,\, +\,\,\,[A(n_{i-1})\bar{A}(n_{i})\widehat{A}(n_{i+1})-\bar{A}(n_{i-1})\widehat{A}(n_{i})A(n_{i+1})]. 
\label{eq:cancellation_scheme_nonequilibrium}
\eea
It is easy to check  that  this   form of $\mathbf{F}(n_{i-1},n_i,n_{i+1})$   indeed  serves the purpose.  Note that,  unlike the 
previous cases  where   we had  only one   kind of  auxiliary  matrix $\widetilde{A}(n),$  here   we have used  three different 
 auxiliary  matrices  $\widetilde{A}(n),\widehat{A}(n),\bar{A}(n)$. In fact,  if   all  three auxiliaries were same i.e. 
 $\widetilde{A}(n)=\widehat{A}(n)=\bar{A}(n),$ then (\ref{eq:cancellation_scheme_nonequilibrium}) reduces to the familiar cancellation 
 scheme studied here in (\ref{eq:cancellation_scheme}) with $R_l=R_r=1$ and correspondingly one obtains a matrix product steady state for 
 totally asymmetric  hoping model  with  hop rate $u_R=u(n_{i-1},n_i,n_{i+1})$ and  $u_L=0,$  a model which  we have   already  discussed 
 in the previous section.

To proceed  further with the   asymmetric hopping model we need to be   more specific  about the dynamics, 
that is, we must be specific about functional forms of $u(.)$ and $v(.).$   
If we consider the functions $u(.), v(.)$ as
\bea
u(n_{i-1},n_i,n_{i+1})=\frac{\langle \alpha(n_{i-1})\mid \beta (n_i-1) \rangle \langle 
\alpha(n_i-1)\mid \beta (n_{i+1}) \rangle }{\langle \alpha(n_{i-1})\mid \beta (n_i) \rangle \langle 
\alpha(n_i)\mid \beta (n_{i+1}) \rangle} \cr
v(n_{i-1},n_{i},n_{i+1})=\frac{\langle \alpha(n_{i-1})\mid \beta (n_{i}+1) \rangle \langle 
\alpha(n_{i}+1)\mid \beta (n_{i+1}) \rangle }{\langle \alpha(n_{i-1})\mid \beta (n_{i}) \rangle \langle 
\alpha(n_{i})\mid \beta (n_{i+1}) \rangle}, 
\label{eq:rate_comps_nonequilibrium}
\eea
then, Eq. (\ref{eq:cancellation_scheme_nonequilibrium}) results in the following solution: 
\bea
\widetilde{A}(n)&=&A(n-1); \bar{A}(n)=A(n+1); \widehat{A}(n)= \theta(n) A(n) \cr
A(n)&=&|\beta(n)\rangle \langle \alpha(n)|,
\label{eq:matrices_asym_diff_ex1}
\eea
where $\theta(n)$  is the  Heaviside step function. 

So, to summarize, if particles on a one dimensional periodic lattice undergo asymmetric hopping with different right and 
left rate functions (constructed below by substituting Eq. (\ref{eq:rate_comps_nonequilibrium}) in (\ref{eq:asym_diff_1}))
\bea
~~~~~~~u_L(n_{i-2},n_{i-1},n_i)\,\,&=&\,\,\frac{\langle \alpha(n_{i-2})\mid \beta (n_{i-1}+1) \rangle \langle 
\alpha(n_{i-1}+1)\mid \beta (n_{i}) \rangle }{\langle \alpha(n_{i-2})\mid \beta (n_{i-1}) \rangle \langle 
\alpha(n_{i-1})\mid \beta (n_{i}) \rangle}\cr
u_R(n_{i-1},n_i,n_{i+1},n_{i+2})\,&=&
\,\frac{\langle \alpha(n_{i-1})\mid \beta (n_i-1) \rangle \langle 
\alpha(n_i-1)\mid \beta (n_{i+1}) \rangle }{\langle \alpha(n_{i-1})\mid \beta (n_i) \rangle \langle 
\alpha(n_i)\mid \beta (n_{i+1}) \rangle},\cr 
&+& u_L(n_{i},n_{i+1},n_{i+2})
\label{eq:rates_asym_diff_1}
\eea
along with $u_R(x,0,z,w)=0$ and $u_L(x,y,0)=0$, the steady state of the  model has a  matrix product form 
$P(\{n_i\}) \sim  Tr \left[\prod_i  A(n_i)\right] \delta (\sum _i n_i -N)$  with matrices 
$A(n)=|\beta(n)\rangle \langle \alpha(n)|$ and the auxiliary matrices  $\widetilde{A}(.)$, $\widehat{A}(.)$ and $\bar{A}(.)$ 
given by  Eq. (\ref{eq:matrices_asym_diff_ex1}).

We  conclude  this subsection with the following remark.  Matrices  $A(n)$ we obtain for  the  asymmetric hopping dynamics 
(\ref{eq:rates_asym_diff_1})  are   same as  those we obtain  for dynamics (\ref{eq:rate_PFSS}). The  auxiliary   
matrices  in two cases   are different, but they do not explicitly appear in the   steady state weights. This 
indicates that  these  two   very different dynamics  lead  to the same  steady state  measure.

\subsubsection{Example 2 :}
In this example we study an asymmetric finite range process where $R_l=R_r=R'_l=R'_r=1$.  In details, we  consider a one dimensional 
periodic lattice with $L$ sites with each site $i$ containing $n_i(\geq 0)$ particles and a particle from a randomly chosen site $i$ 
(if not vacant) jumps either to its right neighbor $(i+1)$ with a hop rate $u_R(n_{i-1},n_i,n_{i+1})$ or to its left neighbor $(i-1)$ 
with rate  $u_L(n_{i-1},n_i,n_{i+1})$. In this    model   both the right and left rate functions are symmetric with respect to the  
the departure site $(i).$ Let us   assume that the steady state probability of any configuration $\{n_i\}$ of this stochastic process 
can be expressed as a product of matrices in the form $P( \{n_i\}) \sim  Tr (\prod_i  A(n_i)) \delta (\sum _i n_i -N)$ where $A(n_i)$ is 
the site occupation matrix corresponding to site $i$ containing $n_i$ particles. The steady state Master equation for this interacting 
particle system reads as 
\bea
  \sum_{i=1}^L  [u_R(n_{i-1},n_{i},n_{i+1})+ u_L(n_{i-1},n_{i},n_{i+1})] Tr[\dots A(n_{i-1})A(n_{i})A(n_{i+1})\dots ] \cr
  -\sum_{i=1}^L[ u_R(n_{i-2},n_{i-1}+1,n_{i}-1) Tr[\dots A(n_{i-2})A(n_{i-1}+1)A(n_{i}-1)\dots]\cr
  +u_L(n_{i}-1,n_{i+1}+1,n_{i+2}) Tr[\dots A(n_{i}-1)A(n_{i+1}+1)A(n_{i+2})\dots]] ~ = ~ 0 \cr \nonumber
\eea
Shifting the sum indexes in the above equation and rearranging them suitably, we arrive at
$\sum_{i=1}^{L} \mathrm{Tr}[ \dots A(n_{i-2})~\mathbf{F}(n_{i-1},n_{i},n_{i+1})~ A(n_{i+2})\dots ]\,=\,0,$ where  
\bea
 \hspace*{-2cm} \mathbf{F}(n_{i-1},n_{i},n_{i+1})\,=\, \left[u_R(n_{i-1},n_{i},n_{i+1})+ u_L(n_{i-1},n_{i},n_{i+1})\right] A(n_{i-1})A(n_{i})A(n_{i+1})\cr
~~~~~~~~~ - u_R(n_{i-1},n_{i}+1,n_{i+1}-1)~ A(n_{i-1})A(n_{i}+1)A(n_{i+1}-1)\cr~~~~~~~~~ - u_L(n_{i-1}-1,n_{i}+1,n_{i+1})~ A(n_{i-1}-1)A(n_{i}+1)A(n_{i+1}).
  \label{eq:master_eq_asym_hop_2}
\eea
So, just like the previous example, the Master equation in steady state has been written as a sum of $L$ terms each containing a three 
site function $F(x,y,z),$  which we must   write in a   way   using auxiliaries  so  that   the terms  within the sum   cancel  with each 
other.  To this end, we   further  specify the   rate functions $u_{R,L}(.)$ as 
\bea
u_R(n_{i-1},n_i,n_{i+1})=\gamma~ \frac{\langle \alpha(n_{i-1})\mid \beta (n_i-1) \rangle \langle 
\alpha(n_i-1)\mid \beta (n_{i+1}) \rangle }{\langle \alpha(n_{i-1})\mid \beta (n_i) \rangle \langle 
\alpha(n_i)\mid \beta (n_{i+1}) \rangle} \cr 
~~~~~~~~~~~~~~~~~~~~ + ~ \delta~ \frac{\langle \alpha(n_{i-1})\mid \beta (n_i-1) \rangle}
{\langle \alpha(n_{i-1})\mid \beta (n_i) \rangle}\langle \alpha(n_{i}-1)\mid \beta (n_{i+1}+1) \rangle \cr
u_L(n_{i-1},n_{i},n_{i+1})=\delta~ \langle \alpha(n_{i-1}+1)\mid \beta (n_{i}-1) \rangle 
\frac{\langle \alpha(n_{i}-1)\mid \beta (n_{i+1}) \rangle}{\langle \alpha(n_{i})\mid \beta (n_{i+1}) \rangle}.
\label{eq:rates_asym_hop_2}
\eea
These hop rates resemble the rate functions considered by the authors in \cite{AZRP_AFRP} in context of asymmetric finite range process.  
Here  too, we use  three   auxiliary matrices  $\widetilde{A}, \widehat{A}$ and $\bar A,$  but now the   last two auxiliary matrices  are  
functions of  two arguments whereas  $\widetilde{A}$  has one argument as in earlier cases.  Explicitly,  
the   cancellation scheme  reads as, 
\bea
  \hspace*{-1.5 cm}\mathbf{F}(n_{i-1},n_{i},n_{i+1}) \, =\,[A(n_{i-1})\widetilde{A}(n_{i})A(n_{i+1})-A(n_{i-1})A(n_{i})\widetilde{A}(n_{i+1})]\cr
\hspace*{-1.5 cm}\qquad \qquad \qquad \,\,\,\,\, +\,\,\,[A(n_{i-1})\widehat{A}(n_i,n_{i+1})A(n_{i+1})-\widehat{A}(n_{i-1},n_i)A(n_i)A(n_{i+1})]\cr
\hspace*{-1.5 cm}\qquad \qquad \qquad \,\,\,\,\, +\,\,\,[A(n_{i-1})\bar{A}(n_{i-1},n_i)A(n_{i+1})-A(n_{i-1})A(n_{i})\bar{A}(n_i,n_{i+1})].
\label{eq:cancellation_scheme_asym_hop_2}
\eea
One can easily check that  Eq. (\ref{eq:cancellation_scheme_asym_hop_2})  satisfies the steady state condition (\ref{eq:master_eq_asym_hop_2}) 
and it results in a matrix product state with   matrices $A(n)$  in the  familiar   form
\be
A(n)~=~|\beta(n)\rangle \langle \alpha(n)|.
\label{eq:matrices_asym_hop_2}
\ee
The corresponding choice of auxiliary matrices  are then 
\bea
\widetilde{A}(n)=\gamma~ A(n-1), ~~~ \widehat{A}(m,n)=\delta~ A(m-1)|\beta(n+1)\rangle \langle \alpha(m)|, \cr
~~~~~~~~~\bar{A}(m,n)=\delta ~|\beta(n)\rangle \langle \alpha(m+1)|A(n-1). 
\label{eq:auxiliary_matrices_asym_hop_2}
\eea
So, if we have an asymmetric particle transfer process with right and left rate functions expressed by (\ref{eq:rates_asym_hop_2})
we have  a  matrix product  steady state, same as the one obtained for dynamics (\ref{eq:rates_asym_diff_1})  or for  (\ref{eq:rate_PFSS}). 

However, it should be mentioned that the cancellation scheme used here in Eq. (\ref{eq:cancellation_scheme_asym_hop_2}) 
is again very much distinct from the schemes used in the previous examples. 

\section{Conclusion}
\label{sec 5.}
We have introduced a matrix product ansatz for  systems  of interacting particles without any hardcore constraints. 
In these class of models   particles on a  one dimensional lattice   jump to their  neighboring sites with  some   rate that 
depends  on the  occupation  of the departure  site and  its neighbors   within a specified range. In  case of MPA for exclusion processes, 
where   particles obey hard core constraints,  we need only a finite number of  matrices to represent each  species of particle. For 
systems without hardcore constraints, the sites can either    be vacant or occupied by   arbitrary number of particles  and thus  a 
matrix product  state   that  describe these systems would   require infinite number  of  matrices (in contrast to the hardcore exclusion 
processes), each  corresponding to a  specific occupation number. Further,  any given   dynamics would insist   the matrices   to   follow  
an  {\it algebra}, consisting of infinitely many matrix-relations.  Finding specific representation of  these  infinite set of matrices  that 
follow the algebra,appears to be   complex, but  here, in this article,  for  a  generic class of models, we    show that the matrices  can 
be  parametrized by the occupation  number (which essentially leads to the name {\it site occupancy matrices} of the matrices $A(n)$),  
 i.e., the elements of the matrix are functions of  the occupation   number. This parametrization actually helps to treat the infinite set 
 of matrix algebra as a single equation of the matrix function $A(n)$ -which can be solved once and for all for any general $n,$ so that 
 one no more has to solve for the matrices $A(0), A(1), A(2)\dots$ separately.

The class   of   hopping  models   we studied here   is very general;  many well known  models, like 
zero range process, misanthrope process,  models  with pair factorized steady state,  and  finite range processes 
are   only    some of the special cases, for which  the  exact steady state  weights  are already known.   
In this article, first we   re-derive the steady state    weights of these models  using matrix product formulation.  

We  also study FRP for very general rates which has not been  studied earlier,  and show that  their  steady state 
can be  expressed  in   matrix product form. A specific example   is   FRP  with $R=2,$   which    leads to a 
3-cluster  factorized steady state   with  weights  $P(\{n_i\}) \sim \prod_{i=1}^L  g(n_{i-1}, n_i, n_{i+1})$ when the
hop-rates satisfy a specific condition. Even  when the steady state  is  known exactly, there  are practical difficulties  in calculating 
the  partition function  or  average steady state values of the observables; this is  because   any particular occupation variable $n_i$  
appears thrice in  the product  and  carrying out  sum  of $n_i$ for all  possible  values  is  non-trivial.  For some special cases, like 
when the   weight function  has a sum form  $g(k,l,m)=  f_0(k) + f_1(l) + f_2(m)$  one  can  write   the steady state in a  matrix product 
form, where matrices   depend on only a  single  occupation  variable $n_i$  which enables us  to carry out the corresponding 
sum over  $n_i.$  Such a matrix product solution has been known for totally asymmetric finite range process \cite{FRP}. In  \cite{FRP}, 
however, re-writing     the  3-cluster factorized steady state  in
a  matrix    product  form   was  only  a   mathematical trick,   a   relation  between  the matrices and  dynamics  of the system were not 
established. When   $g(k,l,m)$   has a `sum-form',   the  matrix product ansatz  formulated here leads  to a  
matrix algebra  which is   naturally satisfied by the matrices constructed  in  \cite{FRP}. Moreover  we  explicitly 
derive  matrix representations for   certain other class of    weight functions $g(k,l,m) = f_0(k)f_1(l)+f_2(l)f_3(m)$  and more generally 
for $g(l,m,n)= \langle f_0(l) |f_1(m) \rangle +\langle f_2(m)|f_3(n)\ra.$
In general there are   no  well defined  methods   to  obtain matrix representation from a  given matrix algebra. Fortunately 
for systems   having a  cluster  factorized  steady state, the  matrix representations that describe a matrix product state can be  
derived  systematically. In the  Appendix  we have   discussed  this  in details.

We  further  study  asymmetric  finite  range processes where  the  rate functions   for right and left hops  are different in the sense 
that  they have  different number of arguments  and/or   different functional  forms. In particular,  we   introduce a model where the  
hop rate for right move  $u_R(.)$ depends on occupation of  departure sites, $R_l$ neighbors to  its left and $R_r$  neighbors to  the right. 
Whereas the left hop rate  $u_L(.)$ depends on  the  departure site  and   $R'_{l,r}$  sites to  its left and right  respectively.  We 
obtain a  matrix product steady state for  two  specific cases (i) $R_l=1\neq R'_l=2$ and $R_r=2\neq R'_r=0$,   (ii) $R_l=R_r=R'_l=R'_r=1$.
Interestingly,  both  models   lead to   same   MPS,  but   the   auxiliaries,  used   in the  cancellation scheme  
to   satisfy   the Master equation in steady state, turns out to be  very different.

There are many other interesting directions to pursue   in  the  study of  matrix product  formulation for  interacting 
particles  in absence  of   hardcore constraints.  One  important     direction  is to  investigate    the open systems, 
where particles  can  enter from left boundary  and  exit from right  boundary of the system.  
It  is well known that open   exclusion processes (EP), where  particles  obey hardcore constraints,  give  rise to 
interesting  results; even the simplest case, namely  totally asymmetric simple exclusion process (TASEP)    
which is exactly solved through  MPA \cite{derrida__tasep_mpa},  shows rich  variety of phases  and transitions    among them as the  
entry and exit rate of particles are varied.   One can   also study   exclusion  processes that   can be mapped  to a 
particular  finite range process.   It is  well  known that,  steady state weight of  exclusion processes can always  be    written  
in matrix product form   if they  can be mapped to   zero-range process \cite{corr_pp_ep}; in this situation   explicit representations 
can be obtained  from the known   steady state weights of the corresponding  zero  range process, which  helps in finding spatial 
correlation  in  EP.  In a similar fashion, using matrix product formulation, one can study the spatial correlation   functions in 
exclusion processes which  can be mapped to finite range processes.

\section*{Appendix:  Matrix product  form  of  cluster factorized steady states}   
  We have  seen in  sections  3. and 4. that the  matrix  product ansatz  naturally  leads to a cluster factorized steady state 
if  the  dynamics of the system  allows one. Depending on the dynamics of the  model, MPA  results in a  specific   
matrix-algebra,   but there are no systematic methods to obtain matrix representation from a  given  algebra.
Thus for models that   has  a  cluster factorized steady state,   it is useful  to     construct the matrices  from  
the known steady state, whenever possible.  We  must remind that,  for FRP with $R \ge2,$   calculating the  partition function 
or  average value of observables    is  not straightforward even when the  exact steady state weights are known in cluster factorized 
form; in such situations   the  matrix  formulation is  certainly a relief.

 To this end we construct the  matrices   for a 3-cluster factorized steady state; it is   straight forward to generalize 
 this for  larger clusters.  Let us consider a specific  CFSS,  $P(\{n_i\}) \sim  \prod_i g(  n_{i-1}, n_{i}, n_{i+1})$  with   
 \be
 g(k,l,m)= \langle f_0(k)|f_1(l)\rangle + \langle f_2(l)|f_3(m)\rangle.
 \label{eq:3_cluster_weight_function}
 \ee
 where  $\la f_\nu(n) |= ( h_\nu^1(n), h_\nu^2(n),  h_\nu^2(n), \dots h_\nu^d(n))$   are    $d$-dimensional row-vectors 
and  $ |f_\nu(n) \ra = \la f_\nu(n)|^T$ (here $\nu=0,1,2,3$). 
These cluster weight  functions  can be rewritten as inner product of vectors and matrices, each of which now  depends on 
{\it a single} individual occupation number. More precisely,   
 \be
 g(k,l,m)=\langle\alpha(k)|\Gamma(l)|\beta(m)\rangle,
 \ee
 \bea
  {\rm where }~ \langle\alpha(k)|= \left(   \begin{array}{cc}
                             \langle f_0(k)| &   1  \\
                           \end{array}
\right);~|\beta(m)\rangle=
\left( \begin{array}{c}
                1 \\
                |f_3(m)\rangle \\
               \end{array}
\right)\cr
{\rm and} ~
\Gamma(l) = \left( \begin{array}{cc}
                |f_1(l)\rangle & 0_{d\times d} \\
                0 & \langle f_2(l)| \\
               \end{array}
\right).
\label{eq:decompose_g}
  \eea
Now the steady state  weights   can be  written as 
\bea
P(\{n_i\}) \sim  \prod_i g(  n_{i-1}, n_{i}, n_{i+1})\cr
=\langle\alpha(k)|\Gamma(l)|\beta(m)\rangle \,\,\,  \langle \alpha(l)|\Gamma(m)|\beta(n)\rangle \,\,\, 
\langle\alpha(m)|\Gamma(n)|\beta(p)\rangle \,\,\, \langle\alpha(n)|\Gamma(p)|\beta(q)\rangle \dots
\cr
=Tr\left[\Gamma(l)|\beta(m)\rangle \langle \alpha(l)|\,\,\,\Gamma(m)|\beta(n)\rangle \langle \alpha(m)|\,\,\,
\Gamma(n)|\beta(p)\rangle \langle \alpha(n)|\dots \right] 
\cr
=Tr\left[G(l,m)\,\,\,G(m,n)\,\,\,G(n,p) \dots\right].
\label{eq:eqivalent_2_cluster}
\eea
Thus  we have transformed the  $3-$cluster weight functions to a  matrix product form   with matrices 
$G(l,m) = \Gamma(l)|\beta(m)\rangle \langle \alpha(l)|$ depending on occupancy of  two neighboring sites.
To get the site occupation matrices $A(n)$ which  depend {\it only on a single site occupation number,}  as in the  matrix product  
ansatz   (\ref{eq:MPA}),  we proceed as follows. Since the  outer product of  any  two  vectors  $|b\ra$ and $\la a|$ can be written as  
\be 
|b\ra\la a|   = ( I \otimes  \la a|) (|b\ra \otimes I)
\ee
with   $I$   being  the  identity matrix   of  same dimension as  that of  $|b\ra$ and $\la a|$,  
we  rewrite  $G(l,m)$ as
\be
G(l,m) = \Gamma(l)|\beta(m)\rangle \langle \alpha(l)| = \Gamma(l)   ~ ( I \otimes  \la \alpha(l)|) ~(|\beta(m)\ra \otimes I).
\ee
 Using   this   in Eq. (\ref{eq:eqivalent_2_cluster})  we get
 \bea
 P(\{n_i\}) \sim  Tr[ \prod_i  A(n_i)] \cr ~~~~~ ~{\rm with}~  A(n) = 
 \left( | \beta  (n) \ra \otimes I\right)  \Gamma(n)  \left(  I \otimes \la \alpha(n) | \right)   \label{eq:matrices_nonequilibrium_R>1-II}
 \eea

In the appendix, we have   demonstrated   how to obtain a  matrix product   form   from a  known    3-cluster factorized steady state. 
There  is no particular   difficulty in extending this formulation to   systems  with  larger  cluster factorized steady state (like 
the steady states of FRP \cite{FRP} with $R>2$).


\section*{References}


\begin{thebibliography}{99}
\bibitem{DDS} {\it Statistical Mechanics of
Driven Diffusive Systems,} Schmittmann, B. and  Zia, R. K. P., 1995  ed. Domb, C. and  Lebowitz, J. L.,  Academic Press, New York.

\bibitem{book1} 
 {\it Nonequilibrium Statistical Mechanics In One Dimension}, ed.  Privman, V., 1997, Cambridge University Press, New York.

\bibitem{book2}  {\it Nonequilibrium Phase Transitions}, Hankel, M.,   Hinrichsen, H., and  L\"ubeck, S, 2009, Springer,  Netherlands.


\bibitem{book_db_1}
 {\it The Principles of Statistical Mechanics}, Tolman, R. C., 1938, Oxford University Press, London, UK.

\bibitem{book_db_2}
 {\it Lectures on gas theory}, Boltzmann, L.,  1964, Berkeley, CA, USA: U. of California Press.

\bibitem{corr_noneq} Karimipour, V.,  Europhys. Lett. \textbf{47} (3), 304 (1999).

\bibitem{multi-species} Basu, U., and  Mohanty, P. K., Physical Review E \textbf{82} (4), 041117 (2010).

\bibitem{corr_pp_ep}
Basu, U., and Mohanty, P. K., J. Stat. Mech. L03006 (2010).

\bibitem{AZRP_AFRP}
Chatterjee, A. K., and Mohanty, P.K.,  arXiv:1704.05386 (2017).


 \bibitem{current_fluctuations}
 Lazarescu, A., J. Phys. A: Math. Theor. \textbf{48}, 503001 (2015).



 
\bibitem{phase_tran_1}
  {\it Phase transitions and critical phenomena}, Domb, C., and  Lebowitz, J.L.,  1972, Academic Press, New York.
 
 \bibitem{phase_tran_2}  Evans, M. R.,  Kafri, Y.,  Koduvely, H. M.,  and Mukamel, D., Phys. Rev. Lett. \textbf{80}, 425 (1998).

\bibitem{book3}  {\it Nonequilibrium Phase Transitions in Lattice Models,}  Marro, J.,  and  Dickman, R., 1999, 
Cambridge University Press, New York.

\bibitem{phase_tran_3}
 Evans, M. R.,  Braz. J. Phys., \textbf{30}, 42 (2000).
 
 \bibitem{phase_tran_4}
 Majumdar, S. N., Krishnamurthy, S. and Barma, M., J. Stat. Phys., \textbf{99}, 1 (2000). 
 

\bibitem{phase_tran_5}  Evans, M. R.,  Levine, E.,  Mohanty, P. K.,  and Mukamel, D., Eur. Phys. J. B  \textbf{41}, 223 (2004).
 
 
\bibitem{bethe1} Faddeev, L. D., arXiv:hep-th/9605187 (1996).

\bibitem{bethe2} Karbach, M.,  and  M\"uller, G.,  Comp. in Phys. \textbf{11}, 36 (1997); Karbach, M., Hu., K., and  M\"uller, G.,  
Comp. in Phys. \textbf{12}, 565 (1998); Karbach, M., Hu., K., and  M\"uller, G., arXiv:cond-mat/0008018 (2000).


\bibitem{bethe3} Mallick, K., and Golinelli., O., J. Phys. A: Math. Gen. \textbf{39}, 12679 (2006).

 
 \bibitem{derrida__tasep_mpa}
 Derrida, B. , Evans, M. R. , Hakim, V.  and Pasquier, V.,  J. Phys A: Math. Gen. \textbf{26}, 1493 (1993).
 
 \bibitem{transfer_mat} {\it Exactly solved models in statistical mechanics,} Baxter, R. J., 1982,  Academic Press Limited, London.
 
 \bibitem{ldf} Touchette, H., Phys. Rep. \textbf{478}, 1 (2009).
 
\bibitem{open_TASEP1} Krug, J., Phys. Rev. Lett. \textbf{67}, 1882 (1991). 
 
 \bibitem{open_TASEP2} Derrida, B., Domany, E., and Mukamel, D., J. Stat. Phys \textbf{69}, 667 (1992).
 
\bibitem{gen_tasep_1}
 Mallick, K.,  Mallick, S. and  Rajewski, N., J. Phys. A \textbf{32}, 8399 (1999).

\bibitem{gen_tasep_2}
 Alcaraz, F. C. and  Bariev,R. Z.,  Braz. J. Phys. \textbf{30}, 655 (2000).

\bibitem{gen_tasep_3}
Hilhorst, H. J., and Appert-Rolland, C.,   J. Stat. Mech. P06009 (2012).

\bibitem{gen_tasep_4}
 Pronina, E., and Kolomeisky, A. B.,  J. Phys. A \textbf{37}, 9907 (2004).
 
 \bibitem{gen_tasep_5}
 Mitsudo, T. and Hayakawa, H., J. Phys. A: Math. Gen. \textbf{38}, 3087 (2005).

 \bibitem{gen_tasep_6}
 Frey, E.,   Parmeggiani, A. and  Franosch, T., Genome Inf. \textbf{15}, 46 (2004);  Helbing, D., Rev. Mod. Phys. \textbf{73}, 1067 (2001).

\bibitem{gen_tasep_7}
  Chowdhury, D.,   Schadschneider, A.  and  Nishinari, K., Phys. Life Rev. \textbf{2}, 318 (2005); 
  {\it Traffic and Granular Flow ’07}, ed.  Appert-Rolland, C. et al., 2009, Springer Verlag, Berlin.

\bibitem{gen_tasep_8}
Neri, I., Kern, N. and Parmeggiani, A.,  Phys. Rev. Lett. \textbf{107}, 068702 (2011).
 
\bibitem{evans_mpa}
 Blythe, R. A., and Evans, M. R., J. Phys. A: Math. Theor. \textbf{40}, R333 (2007). 



\bibitem{corr_eo_ep}
Chatterjee, A. K., Daga, B. and Mohanty, P. K., Phys. Rev. E \textbf{94}, 012121 (2016).

\bibitem{algebraic_bethe}
{\it Quantum Inverse Scattering Method and Correlation Functions}, 
Korepin, V. E.,  Bogoliubov, N. M., and  Izergin, A. G., 1993, Cambridge University Press, Cambridge, U.K..

\bibitem{mpa_bethe_markovian_1d}
Golinelli, O. and Mallick, K., J. Phys. A: Math. Gen. \textbf{39}, 10647 (2006).

\bibitem{bethe_mpa_heisenberg_chain}
Katsura, H. and Maruyama, I., J. Phys. A: Math. Theor. \textbf{43}, 175003 (2010).


\bibitem{mpa_sum_corr_ran_var}
Angeletti, F., Bertin, E. and Abry, P.  EPL \textbf{104}, 50009 (2013).

\bibitem{cmpa_var_states}
Verstraete, F. and Cirac, J. I., Phys. Rev. Lett. \textbf{104}, 190405 (2010).

\bibitem{cmpa_bose_gas}
Aruyama, M., and Katsura, H., J. Phys. Soc. Jpn. \textbf{79} (7),  073002 (2010).

\bibitem{cmpa_interacting_fermions}
Chung, S. S., Bauman, S., Sun, K., and Bolech, C. J., J. Phys.: Conf. Ser. \textbf{702}, 012004 (2016).

\bibitem{mpa_quantum_infor}
Perez-Garcia, D.,  Verstraete, F.,  Wolf, M. M., and Cirac, J. I.  Quantum Inf. Comput. \textbf{7},  401 (2007).


\bibitem{ZRP} Spitzer, F., Adv.Math. \textbf{5}, 246 (1970).


\bibitem{ZRPrev} M. R. Evans and  T. Hanney,  J. Phys. A: Math. Gen. \textbf{38}, R195 (2005).


\bibitem{misanthrope}
Cocozza-Thivent, C.,  Z. Wahr. Verw. Gebiete \textbf{70}, 509 (1985).

\bibitem{FRP}
Chatterjee, A., Pradhan, P. and Mohanty, P. K., Phys. Rev. E \textbf{92}, 032103 (2015) .


\bibitem{wealth_con}
Burda, Z., Johnston, D., Jurkiewicz, J., Kaminski, M., Nowak, M. A., Papp, G. and Zahed, I., Phys.Rev.E \textbf{65}, 026102 (2002).

\bibitem{jamming}
Chowdhury, D., Santen, L., and Schadschneider, A., Phys. Rep. \textbf{329}, 199 (2000).


\bibitem{misanthrope_fss}
Evans, M. R. and Waclaw, B., J. Phys. A: Math. Theor. \textbf{47}, 095001 (2014).

\bibitem{pfss}
Evans, M. R., Hanney, T. and Majumdar, S. N., Phys. Rev. Lett. \textbf{97}, 010602 (2006).






\end{thebibliography}
\end{document}